\documentclass[12pt,preprint,usenatbib]{emulateapj}

\usepackage{amssymb}
\usepackage{times}
\usepackage{pifont} 
\usepackage{rotating}

\newcommand{\actaa}{Acta Astron.}
\newcommand{\pasa}{PASA}

\newcommand{\muhz}{$\mu$Hz}
\newcommand{\numax}{$\nu_{\mathrm{max}}$}
\newcommand{\dnu}{$\Delta\nu$}

\newcommand{\kepler}{\textit{Kepler}}

\newcommand{\keplermission}{\textit{Kepler Mission}}

\shorttitle{Non-radial modes in semi-regular variables}
\shortauthors{Stello et al.}

\begin{document}

\title{Non-radial oscillations in M-giant semi-regular variables: Stellar
  models and \textit{KEPLER} observations}
\author{
Dennis~Stello,\altaffilmark{1,2} 
Douglas~L.~Compton,\altaffilmark{1} 
Timothy~R.~Bedding,\altaffilmark{1,2} 
J{\o}rgen~Christensen-Dalsgaard,\altaffilmark{2,3} 
Laszlo~L. Kiss,\altaffilmark{4,1,5}
Hans~Kjeldsen,\altaffilmark{2} 
Beau Bellamy,\altaffilmark{1} 
Rafael~A.~Garc{\'\i}a,\altaffilmark{6} 
Savita~Mathur,\altaffilmark{7} 
}
\altaffiltext{1}{Sydney Institute for Astronomy (SIfA), School of Physics, University of Sydney, NSW 2006, Australia}
\altaffiltext{2}{Stellar Astrophysics Centre, Department of Physics and Astronomy, Aarhus University, Ny Munkegade 120, DK-8000 Aarhus C, Denmark}
\altaffiltext{3}{Kavli Institute for Theoretical Physics, University of California, Santa Barbara, CA 93106, USA}
\altaffiltext{4}{Konkoly Observatory, Research Centre for Astronomy and
Earth Sciences, Hungarian Academy of Sciences, Budapest, Hungary}
\altaffiltext{5}{ELTE Gothard-Lend\"ulet Research Group, Szombathely, Hungary}
\altaffiltext{6}{Laboratoire AIM, CEA/DSM-CNRS, Universit\'e Paris 7 Diderot, IRFU/SAp, Centre de Saclay, 91191, Gif-sur-Yvette, France}
\altaffiltext{7}{Space Science Institute, 4750 Walnut street Suite 205, Boulder, CO 80301, USA}


\begin{abstract}
The success of asteroseismology relies heavily on our ability
to identify the frequency patterns of stellar oscillation modes.
For stars like the Sun this is relatively easy because the mode frequencies
follow a regular pattern described by a well-founded asymptotic relation.
When a solar like star evolves off the main sequence and onto the red
giant branch its structure changes dramatically resulting in changes in the
frequency pattern of the modes.  
We follow the evolution of the adiabatic frequency pattern from the main
sequence to near the tip of the red giant branch for a series of models. 
We find a significant departure from the asymptotic relation for the non-radial
modes near the red giant branch tip, resulting in a triplet frequency
pattern. 
To support our investigation we analyze almost four years of \kepler\ 
data of the most luminous stars in the field (late K and early M type) and
find that their frequency spectra indeed show a triplet pattern dominated
by dipole modes even for the most luminous stars in our sample.
Our identification explains previous results from ground-based observations
reporting fine structure in the Petersen diagram and sub ridges in the
period-luminosity diagram.  
Finally, we find `new ridges' of non-radial modes with frequencies below
the fundamental mode in our model calculations, and we speculate they are
related to f modes. 
\end{abstract}



\section{Introduction} 
At the beginning of this century, the semi-regular variables were
probably the least understood of all variable stars \citep{Wood00}.  In the 
decade following, we have seen tremendous progress in the study of these
stars driven by long-duration ground-based surveys such as MACHO 
\citep{Wood99,Fraser05,Riebel10},
OGLE \citep{KissBedding03,Soszynski04,Groenewegen04,Ita04a,Soszynski07},
EROS \citep{Spano11}, and even `backyard observations' \citep{Tabur10}.  
With four years of continuous data from \kepler, it has now become possible to 
investigate the frequency spectra of late K and early M giants 
using space-based data \citep{Banyai13,Mosser13}.   

Semi-regular variables bridge the gap between the lower luminosity G and K
giants oscillating in multiple radial and non-radial modes like the Sun
\citep{ChaplinMiglio13,Hekker13,Mosser13a,GarciaStello14}, and the
radial fundamental-mode `Mira' oscillators \citep{OlivierWood05}.  
Key uncertainties concerning semi-regular variables include 
the transition between so-called `solar-like' to the 
much  
higher amplitude `Mira-like' oscillations
\citep{Dalsgaard01,Bedding05a,XiongDeng13}.  This includes 
the question of non-radial modes, which are definitely present in less luminous
giants but absent in Miras. 
There has been a long discussion on
the origin, and ultimately the mode orders, associated with the main ridges
seen in the period-luminosity diagram for these giants
\citep[e.g.][]{WoodSebo96,Wood99,BeddingZijlstra98,KissBedding03,Ita04,Tabur10,DziembowskiSoszynski10,SoszynskiWood13,Soszynski13,Takayama13,Mosser13}.   
Throughout this discussion, it has frequently been assumed that
the oscillations in semi-regular variables consist mainly or
entirely of radial modes
\citep[e.g.][]{Wood99,Dziembowski01,Soszynski13}.  It was suggested by
\citet{Dziembowski01} and \citet{Dalsgaard04} that strong radiative damping
through the coupling to the g modes in the core would make 
the non-radial modes unobservable.  However, using OGLE data 
\citet{Soszynski04} noticed features in the so-called Petersen
diagram of period ratios that they could not explain purely by
radial modes. 
\citet{Soszynski07} refined this result, showing fine
structure in the Petersen diagram that
they suggested could be due to non-radial modes.  
In an attempt to match model frequencies to the OGLE data,
\citet{DziembowskiSoszynski10} calculated both radial and dipole mode  
frequencies but found significant discrepancies between models and
observations.   
In addition to a series of radial modes, \citet{Takayama13} derived
frequencies of a dipole and also a quadrupole mode to explain the fine
structure in the Petersen diagram observed by \citet{Soszynski07}.  
With the shorter but much higher photometric quality \kepler\ data, \citet{Mosser13}
clearly detected non-radial modes among the least luminous stars in
their sample.  However, they found that the power of the dipole modes
decreased dramatically for luminous stars having dominant oscillation
frequencies below \numax\ $\sim 1\,$\muhz\ (period $\gtrsim
10\,$days). During this transition the  
power of the radial modes increased. Clear evidence for non-radial modes in
high-luminosity (M-type) red giants has not been established.

In this Letter we examine the oscillations in late K and early M giants
using both theoretical adiabatic frequency calculations of
full non-truncated stellar models and four-year \kepler\ data of about two
hundred stars.
We show that these stars oscillate in both radial and non-radial modes but
with frequency spectra that differ markedly from those of less luminous giants.  
These differences explain the fine structure seen in the Petersen diagram.
Most remarkably, we show that
the frequency spectra of M giants are actually dominated by dipole ($l=1$)
modes.

\section{Theoretical models and adiabatic frequencies} \label{theory}
Calculating frequencies of non-radial oscillations in luminous giants is
challenging and computationally intensive because they are mixed modes that
arise from coupling between p modes in the envelope and a large
number of extremely high-order g modes in the core.
Previously, the presence of the core and the coupling with the g modes were
neglected \citep[e.g.][]{DziembowskiSoszynski10} or dealt with using a
non-adiabatic asymptotic treatment \citep{Dziembowski12}.
Our frequency calculations are based on non-truncated models and 
take the coupling between envelope and core fully into account. 
This minimizes the risk of inducing sudden artificial frequency
shifts from truncation and hence allows us to follow the oscillation modes
from the main sequence to 
the red giant branch with a self consistent set of model calculations. 
We calculated adiabatic frequencies using an updated version of ADIPLS
\citep[][; Christensen-Dalsgaard et al., in preparation]{DalsgaardAdipls08}. 
To represent the 
high-order g-dominated mixed modes in the core, the spatial mesh was distributed
according to the asymptotic behavior of the eigenfunctions, and the scan for
eigenfrequencies was similarly based on the asymptotic distribution. For
the most evolved models the g-mode behavior was inadequately resolved by
the spatial mesh, even at the highest number of points (around 76,000);
however, we have ascertained that this had insignificant effect on the
properties of the acoustically dominated modes that we are concerned with
here. 
Our stellar models are derived using the MESA \texttt{1M\_pre\_ms\_to\_wd}
test suite case \citep{Paxton11,Paxton13}. 
We cross-checked our results using ASTEC models \citep{DalsgaardAstec08} and
frequencies derived using GYRE \citep{TownsendTeitler13} and found no
significant difference to the MESA-ADIPLS results presented here.

\begin{sidewaysfigure*}
\vspace{9cm}
\includegraphics[width=24.5cm]{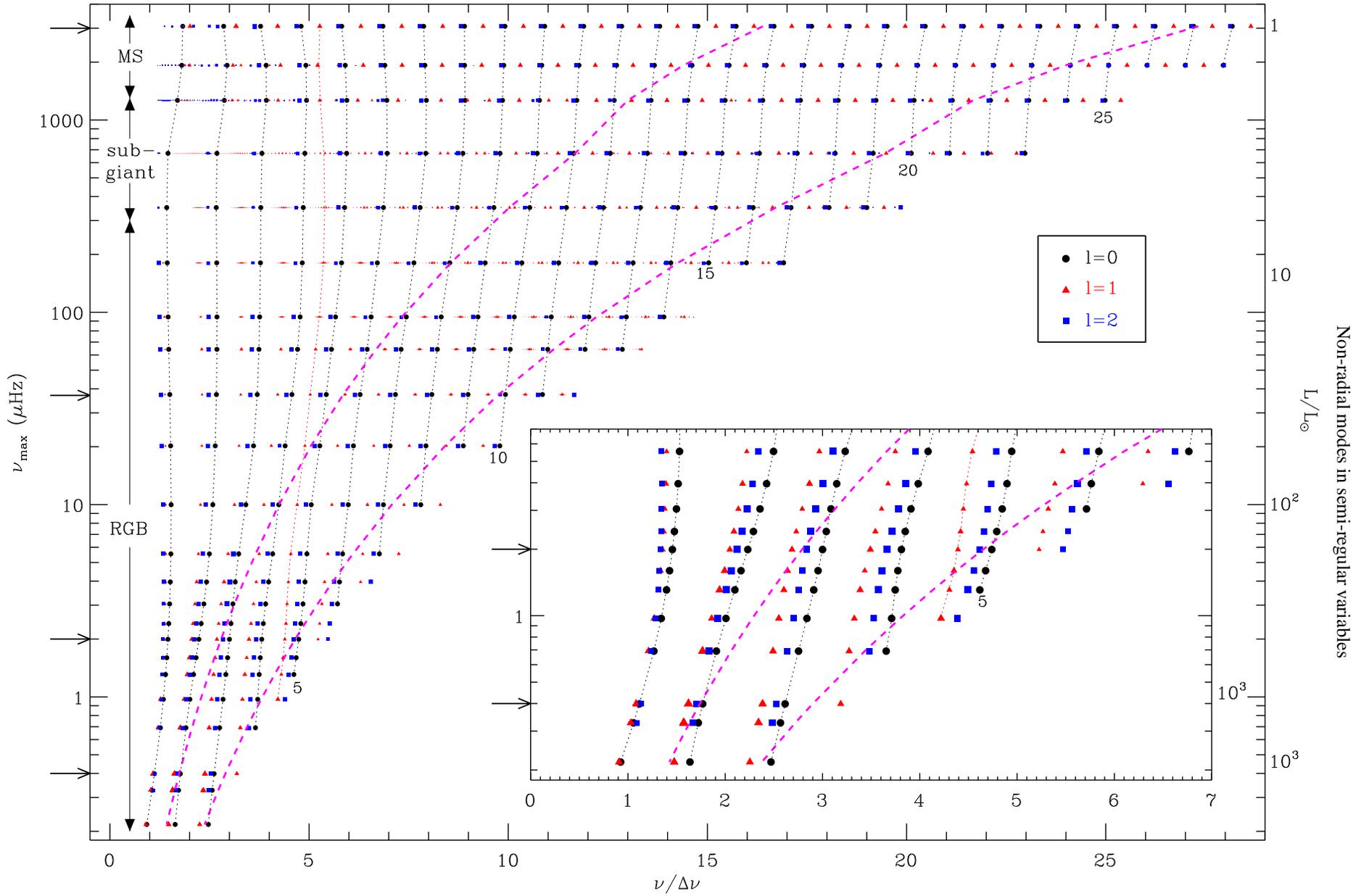}
\caption{Model frequencies, in units of the asymptotic large
  frequency separation, for solar-metallicity models along a 1M$_\odot$ track.
  Each model is plotted according to its \numax.  Symbols for different
  spherical degrees are indicated to the right, and evolutionary states are indicated
  to the left. Dotted vertical lines connect radial (black) and dipole
  (red) modes of the same p-mode
  order, the former annotated for every 5$^{\mathrm{th}}$ order. Dashed magenta lines
  show locations of 75\% and 125\% of \numax.  Horizontal arrows mark models
  shown in Figure~\ref{echelles}. The inset shows a close-up
  of the lower-left region. 
\label{allmodes}} 
\end{sidewaysfigure*}
Figure~\ref{allmodes} shows model frequencies for an evolving star of
1M$_\odot$. Each symbol represents a mode of degree $l=0$ (radial), $l=1$
(dipole), or $l=2$ (quadrupole).    
The symbol sizes are scaled by 
$1/\sqrt{\mathrm{Inertia_{mode}}}$,
normalized to the radial modes, and 
resemble a frequency-dependent `pseudo' amplitude \citep{Aerts10}.  
To guide the eye, we
have connected radial modes of the same order from one model to the next by
dotted black lines.  We show all frequencies from
about 0.1 to 0.9 times the acoustic cut-off frequency.  Assuming the
oscillations are solar-like, the modes between the two dashed lines 
are the strongest and most likely to be observed.  

Between the dashed lines, the main-sequence models show a 
pattern closely resembling the asymptotic relation
\citep{Tassoul80}, with dipole modes 
about halfway between each closely spaced quadrupole-radial pair. 
When the model reaches the subgiant phase, 
the core contraction causes the frequencies of the g modes in the core to
increase while envelope expansion leads to a decrease of the 
p-mode frequencies,
eventually resulting in an overlap of the two sets of modes \citep{Dalsgaard95}.
This leads to coupling of the non-radial modes, resulting in mixed modes 
whose frequencies can become strongly bumped from the regular p-mode
pattern seen during the main-sequence.
The effect first becomes
visible for the quadrupole modes \citep{Stello12}, quickly followed by the
dipole modes, which show the strongest mode bumping (Figure~\ref{allmodes},
models with \numax$\,\lesssim 1500\,$\muhz). 

As the model evolves up the red giant branch the core
keeps contracting, resulting in an increase in the density of g modes and
in the width of the evanescent zone separating the g- and p-mode cavities. 
This gives rise to many more mixed modes and a weaker coupling than in the
subgiant phase, and we start to see clusters of 
$l=1$ mixed modes at each p-mode order
($30\lesssim\,$\numax/\muhz$\,\lesssim 500$)
\citep{Dupret09,Bedding10}.  
This is because several g modes in the core are coupling to each p mode in
the envelope.
The largest symbols within each cluster reveal the location
of the resonant p mode and allow us to follow the
evolution of the spacing between this mode and its nearest radial mode
along the red giant branch.
As pointed out by \citet{Bedding10}, the dipole modes have moved slightly to
the right of centre relative to the radial modes 
\citep[see also][]{Mosser10universal,Huber11,Corsaro12}, a result that was
also seen in a series of stellar models by \citet{Montalban10}.   
In Figure~\ref{allmodes} we see that the dipole modes move even further to
the right for more luminous models, eventually forming the left fork of a
`triplet', with the quadrupole mode in the middle and the radial mode to the
right (see inset).   

The change in the frequency pattern from the
main sequence to the 
luminous giants is due to two factors.  Firstly,
the excitation shifts to lower-order modes (\numax\ decreases), 
away from the asymptotic regime. Secondly, we see a
change in the frequency pattern at 
fixed radial order. For example, the
positions of the dipole modes (red dotted line) clearly shift relative to that 
of the radial $n=5$ ridge. 
\begin{figure*}
\includegraphics[width=18cm]{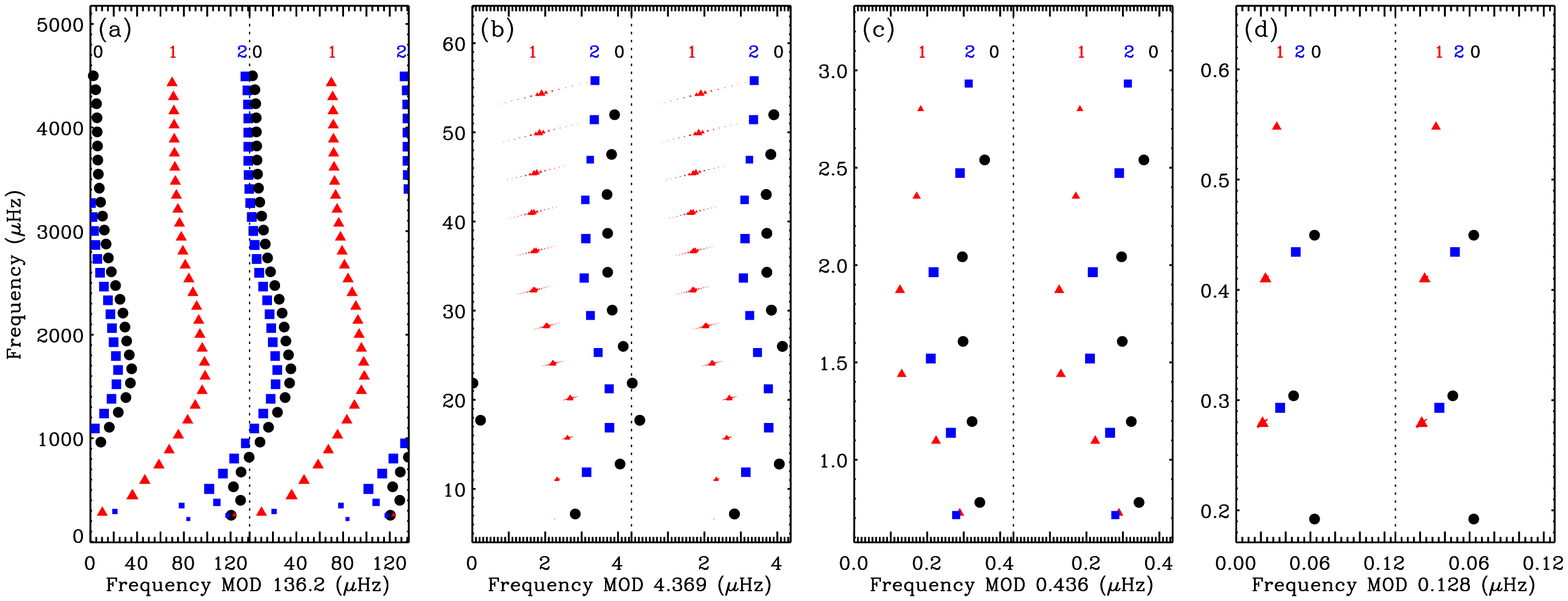}
\caption{\'Echelle diagrams for four representative models from the
  main sequence to near the tip of the red giant branch. Mode degrees are 
  indicated for each ridge, and symbol size is scaled as in
  Figure~\ref{allmodes}. Each \'echelle is plotted twice, as indicated by the
  vertical dotted line.
\label{echelles}} 
\end{figure*} 

The change in frequency pattern is further illustrated in
Figure~\ref{echelles} with
\'echelle diagrams for the 
four 
models marked by arrows in Figure~\ref{allmodes}.
The main-sequence model (Figure~\ref{echelles}(a)) shows dipole modes 
roughly halfway between the radial modes essentially all the way down to
the lowest-order modes.  The next example (Figure~\ref{echelles}(b)) shows an
intermediate pattern, where the highest frequency dipole
modes are also about halfway between the radial modes, but at lower
frequencies they shift closer to the quadrupole-radial pair,
resulting in triplet structures.  Finally, the very
luminous models (Figures~\ref{echelles}(c) and (d)) show triplet structures
for the entire range. 

Another interesting feature in Figure~\ref{allmodes} is the presence of
non-radial modes with frequencies below the fundamental radial mode.
They have low inertia to be visible
in Figure~\ref{allmodes}.  These may be related to the `f modes' found by
\citet{Cowling41} to be intermediate between the
g and p modes, in calculations neglecting the perturbation to the
gravitational potential (the so-called Cowling approximation). Such modes,
with no node in the radial direction, are also found in the Cowling
approximation for realistic unevolved models, with frequencies below the
fundamental radial mode and inertias only slightly higher than that
mode. When the Cowling approximation is not made there is no such f mode
for $l = 1$, since with no radial node it would displace the center of
mass; however, the f mode with $l = 2$ is still found. A similar mode
confined to the envelope could account for the potentially visible $l = 2$
mode below the fundamental radial mode. For the corresponding $l = 1$  mode
found in the calculations, largely trapped in the envelope, we speculate
that it is similar to the f mode in the Cowling approximation but with a
structure in the core ensuring that the center of mass is not
displaced. This calls for further investigation. 
We speculate that these modes might be related to the extra
sequence F in the period-luminosity diagram near the fundamental
sequence C \citep{SoszynskiWood13}.

\section{Observations}
We investigated the oscillation spectra of 195 
giants that were pre-selected as M-giant targets for the \keplermission,
based on their variability in the ground-based ASAS survey
\citep{Pigulski09}\footnote{Our data were selected as part of the KASC
  Working Group 12 activities.}.   
The \kepler\ data were obtained between 
2009 June 20 and 2013 April 4, corresponding to observing quarters 2 to
16, using the spacecraft's long-cadence mode 
($\Delta t \simeq 29.4\,$minutes).  
To correct quarterly flux discontinuities in the data we first grouped stars
into `fast' and `slow' oscillators based on the number of intercepts
of the time series relative to a linear fit to the data using only the first
quarter (Q2). We chose 15 crossings as the discriminator, corresponding to a
dominate oscillation period of about 11 days. 
For the `fast' group we subtracted a linear fit from each quarter and then shifted
their mean flux levels to that of the first quarter.
For the `slow' group we made linear fits to the five points on
either side of quarterly gaps and shifted the flux to make the two fits
intersect at the middle of the gap. Again, we used the first quarter as
anchor point for the flux level. However, where the gap was longer than 10
days, we applied the same method as for the `fast' oscillators.

\begin{figure*}
\includegraphics[width=18.6cm]{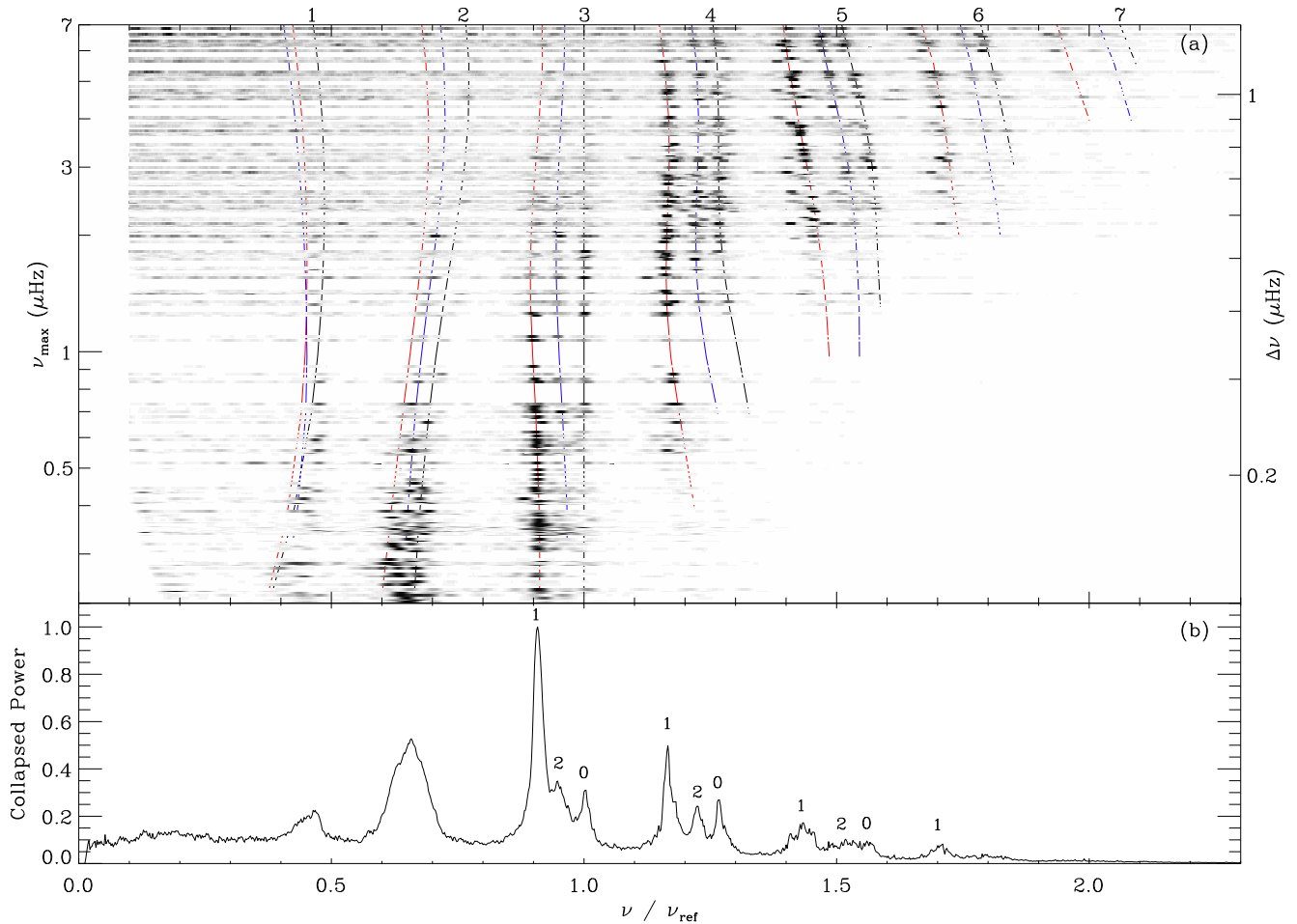}
\caption{(a) Observed power spectra of
  195 \kepler\ stars, each displayed as a horizontal band with
  the level of power indicated by the gray scale.  For
  comparison, ridges of model frequencies are shown in the background for angular degrees
  $l=0$ (black), $l=1$ (red), and $l=2$ (blue).  The radial order
  is indicated at the top, starting from the fundamental radial mode ($n=1$). (b)
  Vertically collapsed
  version of panel (a) with angular degrees indicated.
\label{f2}} 
\end{figure*}

We ordered the stars by \numax\ and stacked their
frequency-scaled power spectra, as shown in Figure~\ref{f2}(a).  We see
clear triplet ridges for each radial order.
To construct Figure~\ref{f2}(a) we used the following two-stage approach.
First, we examined each spectrum in an attempt to determine a
reference frequency, $\nu_\mathrm{ref}$, of a clearly identifiable mode.
For this stage, the reference mode did not need to be the same for all stars.  
To aid the identification we looked for the 
strongest modes, assuming they were the ones closest to \numax, and
compared with the frequency pattern and \numax\ from stellar models
(Figure~\ref{allmodes}).  This allowed us to associate  
the correct radial order with each observed triplet
structure  for about 40\% of the stars.
For each remaining star we made an initial guess of the reference
mode identification and frequency.  We then created a
template spectrum as the average of the 10 already-identified spectra
with reference frequencies closest to that of the initial guess for the
target star.  A comparison 
revealed the correct order of the reference mode for the target star. 

With the modes identified for all stars, we then sought to associate values
of $\nu_\mathrm{ref}\equiv\nu_{n=3,l=0}$, \numax, and \dnu\ for each star based on
matching with a grid of models.  
The main purpose of this second step was to  align the spectra relative
to a common reference mode and to sort the stars by \numax\ with
minimal effect from observational uncertainty.
The initial matching model, and hence the initial \numax, was found
by interpolating the models that bracketed the observed reference frequency
found in the previous step.
The final best-matching model was found by first generating 50 additional
interpolated models that sampled the $\pm20\,$\% range around the initial \numax.
From these models we then created model spectra, where each mode was 
represented by a Lorentzian profile with a width equal to the inverse
observing time, and mode heights were modulated by a Gaussian envelope 
centered on \numax\ with $\mathrm{FWHM} = 0.25\,$\numax. By correlating the
observed and modelled spectra, we identified the final best model as the
one with the strongest correlation.  The resulting \numax\ was used to sort
the stars in Figure~\ref{f2}(a) and $\nu_\mathrm{ref}$ was used to scale the
frequency axis.  
Each spectrum is represented by a horizontal band with the level of power
indicated by the grayscale.  The power was normalized for each spectrum by
its highest peak. The low-frequency noise was suppressed by multiplying the
spectrum by a Gaussian envelope of FWHM = \numax\ centered at \numax. 
For the less-evolved stars, we applied extra smoothing of the spectra
proportional to $\nu_\mathrm{ref}$ in order to produce a common width of the
mode ridges. 
It is quite remarkable how closely the observations follow the model frequencies
(solid lines in the background) in this very non-asymptotic regime. 
This agrees well with the trend seen among K giants, which
shows a decreasing offset between the observed and modelled
frequencies towards more luminous stars \citep{White11}.

In Figure~\ref{f2}(b) we show the collapsed version of panel (a). 
This demonstrates clearly that the dipole modes are
generally much stronger than the radial modes, unlike
in less evolved stars \citep[see also][~and references therein]{Mosser13}.
It is particularly interesting to note that the dipole modes 
dominate even for the most evolved stars in our sample (periods $\sim
50\,$days), which seems to go 
against previous interpretations of semi-regular variables
\citep[e.g.][]{WoodSebo96,Wood99,Soszynski07}. 
We further note that the triplet structure disappears for the lowest order
modes, where we see only a single peak per order. This could
be because the ridges of different degrees merge, as
suggested by the models, which we cannot adequately resolve with the length of the
current data set. 

\begin{figure}
 \includegraphics[width=8.8cm]{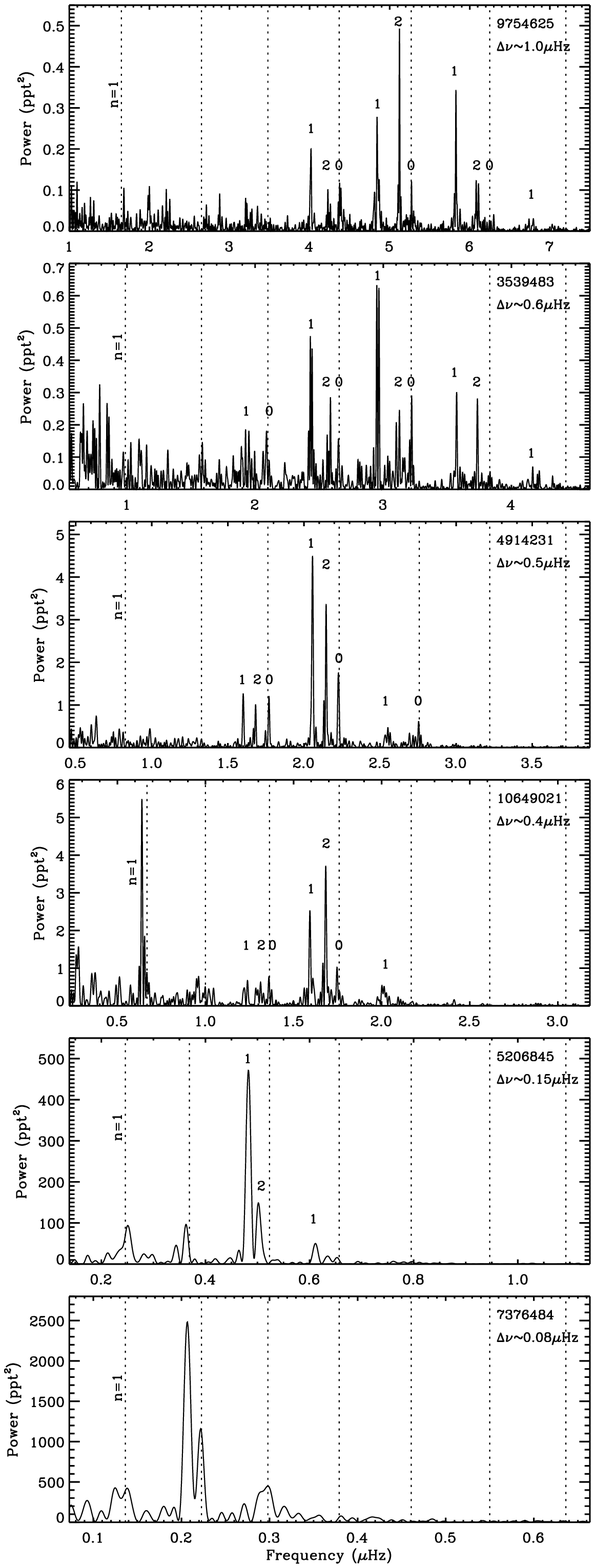}
\vspace{-0.7cm}
\caption{Representative power spectra of stars in our sample with \numax\ ranging
  $\sim 0.2$--5.0\muhz.  The frequency range has been chosen to roughly align the
  radial modes indicated by dotted lines for order $n=\,$1--7
  ($n=1$ being the fundamental) from interpolated best-fitting models (see
  text for details).  The KIC-ID, \dnu, and the mode degree, $l$,
  are shown.    
\label{f1}} 
\end{figure}

Finally, in Figure~\ref{f1} we show a representative selection of power
spectra for five stars. 
Based on the above mode identification we indicate the
most significant non-radial modes and the first seven radial modes 
(the fundamental plus the $1^\mathrm{st}-6^\mathrm{th}$ overtones) from the
best-fitting model found by the interpolation of our stellar model grid.    
A striking feature in some of these spectra is the frequency pattern forming
triplets.

\section{The period-luminosity and Petersen diagrams}\label{petersen}
\begin{figure}
 \includegraphics[width=8.8cm]{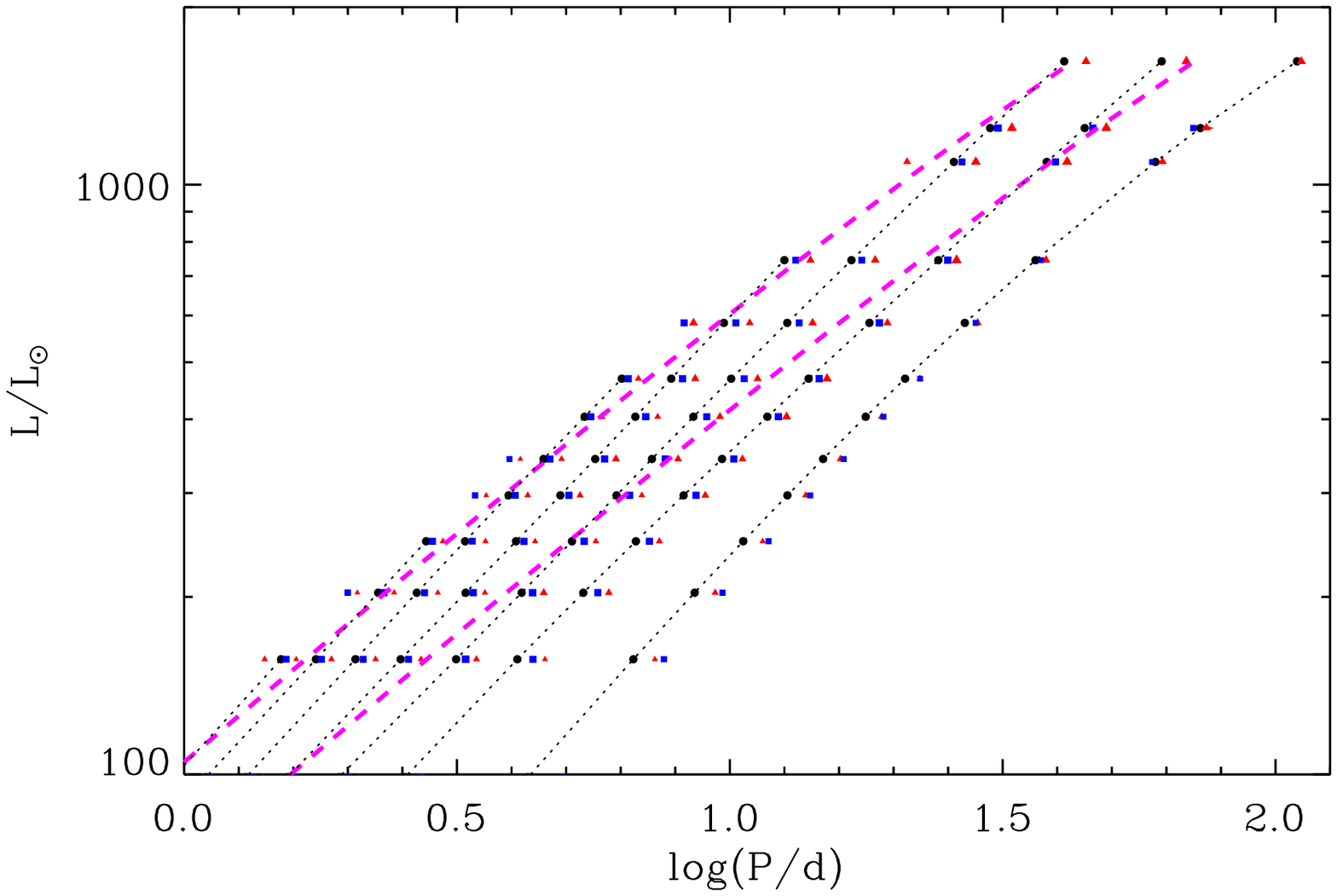}
\caption{Period-Luminosity relation of M giant models of
  1M$_\odot$. Notation follows that of Figure~\ref{allmodes}
\label{pl}} 
\end{figure}

Previous analyses of  frequencies of semi-regular variables have been
based on the period-luminosity diagram and also on the Petersen diagram of
period ratios.  To allow comparison with previous work on such stars, we
show in Figure~\ref{pl} the period-luminosity diagram of the models 
 shown in Figure~\ref{allmodes} (inset) and Figure~\ref{f2}.  

The frequency pattern created by the non-radial
modes seen in Figures~\ref{allmodes} and \ref{echelles} explains
the fine structure in the Petersen diagram seen as sets of
closely-spaced horizontal triplet bands \citep[e.g.][]{Takayama13}, referred
to as sub-ridges in the period-luminosity diagram by \citet{Soszynski07}.
The highest period ratio of $\sim0.98$ mentioned by Takayama et al. (their Figure
7) would emerge from the ratio of two peaks in the power spectrum separated
by the frequency resolution of their data set ($1/T_\mathrm{obs}$), thus
associated with the same mode and presumably arising due to stochastic
excitation of damped modes, as in less luminous giants.  Assuming the
excitation is solar-like, we expect all 
the radial and non-radial modes are excited within a `broad' envelope
centered around \numax.  Due to geometrical cancellation, however, we only
see modes of degree $l \leqslant 2$ (Figure~\ref{f2}), whose period ratios form the
remaining triplet structures.

\section{Conclusions}
Our results clearly show the presence of both radial and non-radial modes in
semi-regular variables.  We demonstrated that the highly non-asymptotic
frequency pattern is dominated by triplets, which is qualitatively
different from that of less luminous 
giants. This explains the fine structure in the Petersen diagram
\citep{Soszynski04,Soszynski07,Takayama13} and, hence, the sub ridges   
in the period-luminosity diagram suggested by \citet{Soszynski07}.

As a general note of caution the few modes excited and the 
divergence from asymptotic behavior of the frequencies strongly suggest
that using measurements of \dnu\ and \numax\ to estimate stellar masses and
radii of luminous giants through the scaling relations could return
dubious results.

Whether the non-radial modes `fade' at 
some point beyond the red giant branch tip, still remains to be seen.
Our observations show that most oscillation power is in the dipole modes
and increasing relative to the radial modes towards the more luminous
stars. This is evidence that non-radial modes do not fade, but our theoretical models
suggest it will be hard to prove because the radial and non-radial modes
merge for the two lowest order modes at high luminosities.  Without decades-long
time series to resolve this, the spectra look like a series of
single-degree overtone modes roughly spaced by \dnu.

Finally, in our model calculations we found non-radial modes of low inertia (hence
visible in Figure~\ref{allmodes}) whose frequencies are below the 
fundamental radial mode. We speculate they may be related to the `f modes'
found by \citet{Cowling41} intermediate between the g and p modes, in
the so-called Cowling approximation.  This intriguing finding calls for
further investigation. 

\acknowledgments
Funding for this Discovery mission is provided by NASA's
Science Mission Directorate. We thank the entire Kepler team
without whom this investigation would not have been possible.
The authors would like to thank Rich Townsend for fruitful discussions and
for implementing the 'JCD' variables option in GYRE.
Funding for the Stellar Astrophysics Centre is provided by The Danish
National Research Foundation (Grant DNRF106). The research is supported by
the ASTERISK project (ASTERoseismic Investigations with SONG and Kepler)
funded by the European Research Council (Grant agreement no.: 267864). 
DS acknowledges support from the Australian Research Council.
LLK has been supported by the MTA Lendület-2009 Fellowship, OTKA Grant
K83790 and the KTIA URKUT 10-1-2011-0019 grant.
SM acknowledges the support by the NASA grant NNX12AE17G and of the
European Community's Seventh Framework Programme (FP7/2007-2013) under
grant agreement no. 269194 (IRSES/ASK).


\end{document}